\documentclass[9pt,twocolumn,twoside]{osajnl}
\usepackage{amsmath,amssymb,amsfonts}
\usepackage{mathrsfs}
\usepackage{graphicx}
\journal{ol}
\setboolean{shortarticle}{true}
\title{Tunable nonlinear spectra of anti-directional couplers}
\author[1,*]{A. Govindarajan}
\author[2]{Boris A. Malomed}
\author[1]{M. Lakshmanan}
\affil[1]{Centre for Nonlinear Dynamics, School of Physics, Bharathidasan University, Tiruchirappalli - 620 024, India}
\affil[2]{Department of Physical Electronics, School of Electrical Engineering, Faculty of Engineering, and the Center for Light-Matter Interaction, Tel Aviv University, 69978 Tel Aviv, Israel}
\affil[*]{Corresponding author: govin.nld@gmail.com}
\ociscodes{(190.0190)   Nonlinear optics; (160.3918)   Metamaterials; (190.3270)   Kerr effect; (060.1810)   Buffers, couplers, routers, switches, and multiplexers; (300.6170)   Spectra.}
\doi{\url{http://dx.doi.org/10.1364/JB.XX.XXXXXX}}
\begin{abstract}
We produce transmission and reflection spectra of the anti-directional coupler (ADC) composed of linearly-coupled
positive- and negative-refractive-index arms, with intrinsic Kerr nonlinearity. Both reflection and transmission feature two highly amplified peaks at two distinct wavelengths in a certain range of values of the gain, making it possible to design a wavelength-selective mode-amplification system. We also predict that a blend of gain and loss in suitable proportions can robustly enhance reflection spectra which are detrimentally affected by the attenuation, in addition to causing red and blue shifts owing to the Kerr effect. In particular, ADC with equal gain and loss coefficients, is considered in necessary detail.
\end{abstract}
\setboolean{displaycopyright}{true}
\begin{document}
\maketitle

Metamaterials, alias left-handed media, are designed to exhibit unusual
light-guiding characteristics, ranging from a negative refractive index
(NRI) to magnetism at optical frequencies, owing to their controllable
dielectric and magnetic properties \cite%
{shalaev2005negative,dolling2006low,zeng2013manipulating,dolling2007negative}%
. While the concept of engineered NRI materials has drawn paramount
interest \cite{NL2012}, to utilize the extraordinary light-guiding
capabilities of metamaterials in applications it is necessary to address
remaining issues \cite{litchinitser2009metamaterials,anantha,lap14,hash17}. Among them, losses
stand out as a major hindrance. Considerable efforts have been invested to
compensate the losses originating from inherent attenuation in metallic
elements of metamaterials, surface irregularities, magnetic resonances, and
quantum effects \cite{dolling2006low,popov2006compensating,popov2007four}.

\begin{figure}[t]
\includegraphics[width=1\linewidth]{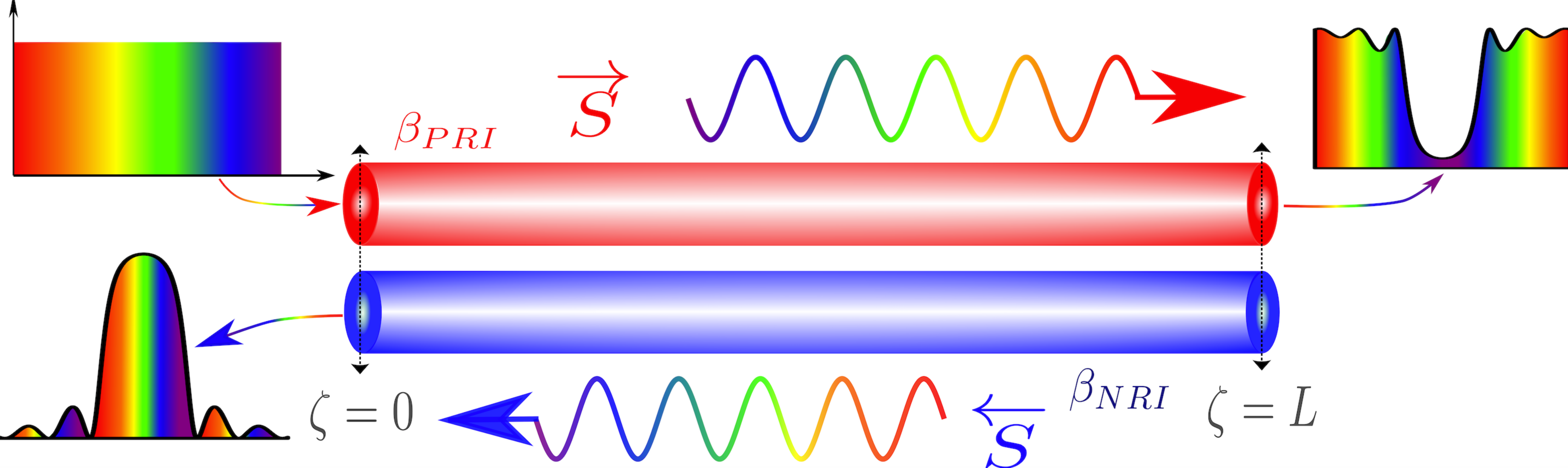}
\caption{A schematic of spectral dynamics in anti-directional couplers
(ADCs) with gain and loss.}
\label{fig0}
\end{figure}

One of the recently developed ramifications of studies of left-handed media is
the design of anti-directional couplers (ADCs), composed of linearly coupled
arms which are fabricated of positive-refractive-index (PRI) and NRI
materials, see Fig. \ref{fig0}. Although the ADC resembles conventional
gain-loss dimers, the contrast between its arms gives rise to light-guiding
characteristics similar to those of distributed feedback structures, such as
Bragg gratings, due to opposite signs of the Poynting vector ($%
\overrightarrow{S}$ in Fig. \ref{fig0}) and phase velocities in the
parallel-coupled NRI and PRI channels \cite%
{litchinitser2007optical,venugopal2011asymmetric,NL2012,govindarajan2019}%
. Accordingly, ADC models exhibit both forward and backward propagation and
optical bi- and multi-stabilities, in addition to the formation of
nonstationary localized modes and discrete solitons in waveguide arrays \cite{zezyulin12}. However, the intrinsic dissipation of the NRI
arm demands the use of large input intensities, which impairs possible
applications to all-optical signal processing \cite{litchinitser2007optical}%
, filters \cite{chu1995}, and amplifiers \cite{malomed1996}.

Recently, Walasik \emph{et al.} have demonstrated that dissimilar
linearly-coupled waveguides with complex refractive-index profiles,
including gain and loss, may maintain meta-$\mathcal{PT}$-symmetric nature
of the coupler, supporting localized modes with real wave numbers \cite%
{walasik2017dissimilar}, and thus extending the concept of stable solitons
in $\mathcal{PT}$-symmetric nonlinear couplers \cite{Driben,Abdullaev,Barash}%
. Also, it has been shown that $\mathcal{PT}$-symmetric dimers can operate
as efficient all-optical soliton switches with low input intensities \cite%
{govindarajan2018tailoring}. Recent work \cite{tennant2019} has laid a
foundation of a new type of lasing systems exhibiting \emph{resonant
amplification} in ADCs . Motivated by those findings, we have demonstrated
that the interplay of independently controlled amplification and attenuation
enables novel bi- and multistable characteristics of ADCs \cite%
{govindarajan2019}. These results suggest that ADCs may exhibit quasi-$%
\mathcal{PT}$-symmetric-like operation, even in the absence of strictly
symmetric and anti-symmetric arrangements of permittivity and gain/loss
profiles. However, to the best of our knowledge no work has reported so far systematic
investigation of nonlinear transmission and reflection spectra of ADCs. We
here focus on this subject, with the intention to analyze the effects of the
gain and loss strengths in the coupled channels on the nonlinear spectra,
and propose feasible applications to all-optical signal-processing networks.

Coupled-mode equations for the light propagation in the nonlinear ADC, with
balanced gain and loss in the PRI and NRI channels, are written as \cite%
{litchinitser2007optical,govindarajan2019}
\begin{gather}
+i\frac{dF(\zeta )}{d\zeta }+\gamma _{1}|F(\zeta )|^{2}F(\zeta )+\kappa
B(\zeta )e^{-i\delta \zeta }=i\chi _{1}F(\zeta ),  \label{eqn:1} \\
-i\frac{dB(\zeta )}{d\zeta }+\gamma _{2}|B(\zeta )|^{2}B(\zeta )+\kappa
F(\zeta )e^{i\delta \zeta }=-i\chi _{2}B(\zeta ),  \label{eqn:2}
\end{gather}%
where $\zeta $ is the longitudinal coordinate, $F(\zeta )$ and $B(\zeta )$
are complex amplitudes of forward- and backward-traveling waves in the PRI
and NRI channels, $\gamma _{1,2}$ are Kerr coefficients in the channels, and
detuning $\delta =\beta _{\mathrm{PRI}}-\beta _{\mathrm{NRI}}\equiv 2\pi $ $%
(n_{F}-n_{B})/\lambda $ is the difference between modal wavenumbers in the
channels, which is determined by modal indices, $n_{F,B}$, and vacuum
wavelength, $\lambda $.  As concerns the
inter-core linear coupling, although its effective coefficients may be
different in the two channels, we adopt equal coefficients, $\kappa $, to
focus on the impact of intra-channel gain ($\chi _{1}$) and loss ($\chi _{2}$%
).
\begin{figure}[tbp]
\centering
\includegraphics[width=1\columnwidth]{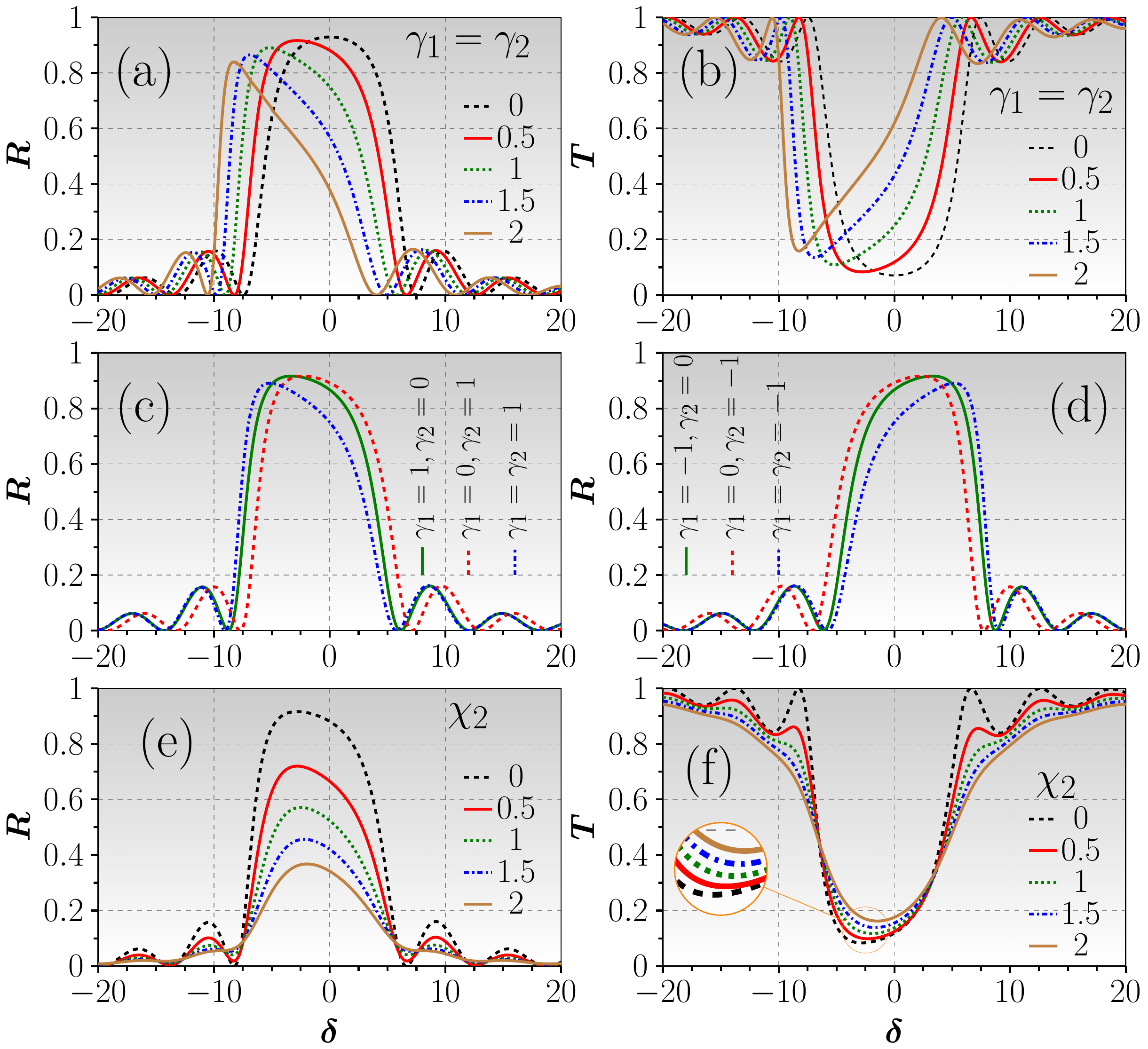}
\caption{Top panels show the nonlinear light dynamics, including the
reflection and transmission, in the conservative symmetric ADC without gain
and loss ($\protect\chi _{1}=\protect\chi _{2}=0$). In the middle panels,
reflection spectra of the asymmetric ADC are shown with (c) self-focusing ($%
\protect\gamma _{1,2}$$\geq $ $0$) and (d) defocusing ($\protect\gamma _{1,2}
$$\leq $ $0$) Kerr nonlinearities. The bottom panels exemplify the dynamics
in the asymmetric nonlinear ADC with a lossy NRI core and the conservative
PRI one ($\protect\chi _{1}=0$), with Kerr coefficients $\protect\gamma _{1}=%
\protect\gamma _{2}=0.5$.}
\label{fig1}
\end{figure}
\begin{figure}[t]
\centering
\includegraphics[width=1\linewidth]{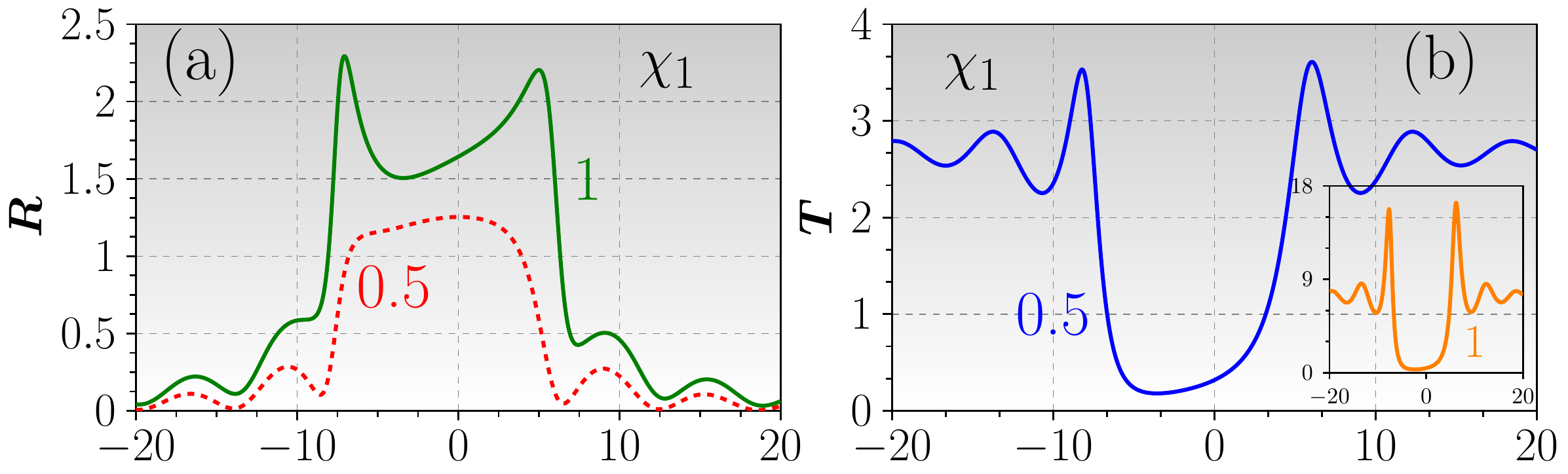}\newline
\includegraphics[width=1\linewidth]{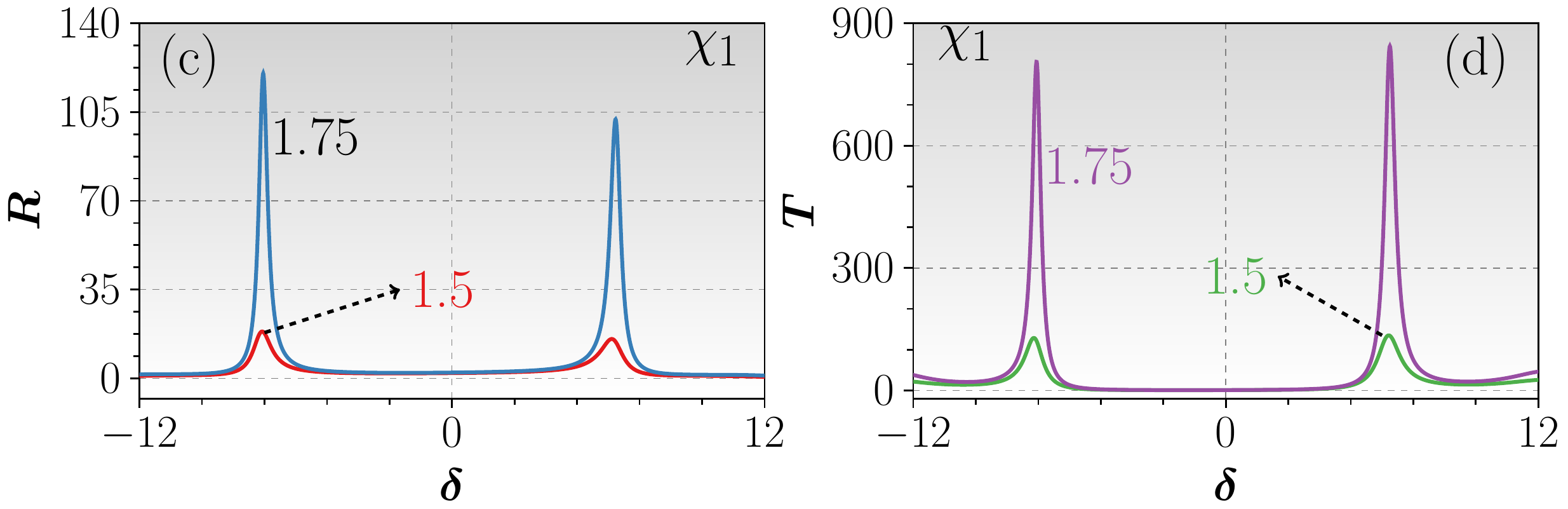} \includegraphics[width=1%
\linewidth]{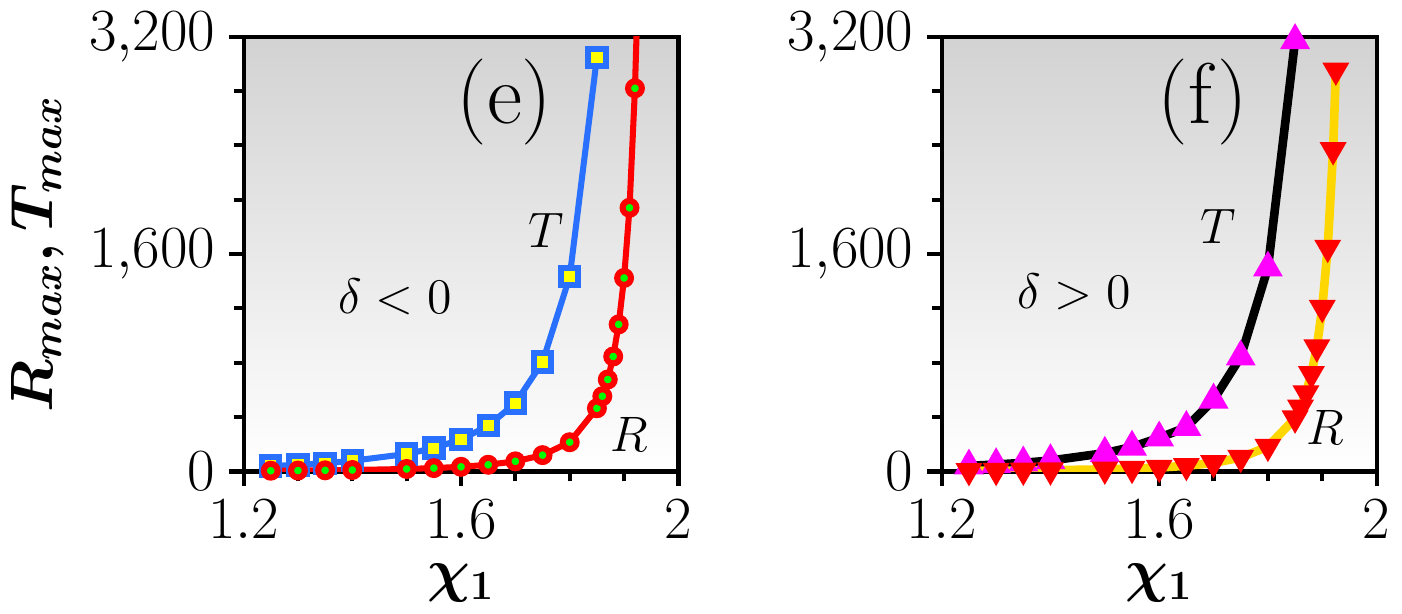}
\caption{Gain-induced ($\chi_1$) mode-selective amplification in the reflection and
transmission spectra of nonlinear ADC with $\protect\gamma _{1}=\protect%
\gamma _{2}=1$, with the gain ($\protect\chi _{1}=1$) applied to the PRI
channel, while the NRI one remains neutral ($\protect\chi _{2}=0$).}
\label{fig2}
\end{figure}

First, we produce nonlinear spectra as a function of detuning $\delta $ (for
linear spectra, one can refer to Ref. \cite{tennant2019}), with a fixed
intensity of the incident signal, $|F(\zeta =0)|^{2}$. To this end, we
integrate the coupled-mode equations (\ref{eqn:1}) and (\ref{eqn:2}) by using the Dormand-Prince method \cite{dormand1978} with appropriate boundary
conditions, keeping scaled values of the system's parameters, including the
ADC's length, $L$, as $\gamma _{1,2}=L=1$ and $\kappa =2$, unless stated otherwise. Then, we define the
transmissivity as $T=|F(\zeta =L)/F(\zeta =0)|^{2}$, where $|F(\zeta =L)|^{2}
$ is the output intensity in the PRI channel. As shown in Fig. \ref{fig1}%
(a), in the ideal (conservative) coupler ($\chi_1=\chi_2=0$), which does not include loss and
gain, the Kerr nonlinearity, with $\gamma _{1}=\gamma _{2}=\gamma $, shifts
the spectral resonance of the reflectivity, $R\equiv |B(\zeta =0)/F(\zeta
=0)|^{2}$, from its position at $\delta =0$ (pertaining to the photonic
bandgap, alias \textit{stopband}, where the spectrum features relatively
high (low) reflection (transmission). Further enhancement of the
nonlinearity results in a farther \textit{red shift} of the spectrum towards
longer wavelengths, with a narrow peak appearing at the top of the
reflectivity spectrum, as seen in Figs. \ref{fig1}(a). Similar peculiarities
are observed in the transmission spectra in the PRI channel, see Fig. \ref%
{fig1}(b). These panels comply with the unitary condition, $T+R=1$, in the
conservative ADC. Unlike the Bragg-grating structures, the light-transfer characteristics
of the ADC is underlain by the transfer of power between the two cores.
Therefore, it is possible to control the transmissivity and reflectivity by
independently adjusting the nonlinearity in the two cores. Accordingly, in
Figs. \ref{fig1}(c,d) we display the spectra for the asymmetric conservative
system, in which either one of the cores (PRI or NRI channel) is linear and
the other one is nonlinear. The plots in Figs. \ref{fig1}(c,d) for the ADC
with the linear NRI or PRI channel (see dashed red and solid blue lines,
respectively) feature a spectral shift towards positive or negative detuning.

The ADC with self-defocusing Kerr nonlinearities ($\gamma _{1,2}<0$)
exhibits opposite trends, \textit{viz}., a \textit{blue shift} of the
spectra towards shorter wavelengths. The respective transmission spectra
reveal features similar to those exhibited in Figs. \ref{fig1}(c,d), and are
not displayed here. All these ramifications clearly indicate that one can
easily tune the ADC spectra with the help of the Kerr nonlinearities in the
NRI and PRI channels. Physically, the origin of the tunability (shifting) 
of the spectra can be understood as follows: the change in the focusing Kerr nonlinear 
coefficient results in an increase of the effective refractive index, which eventually shifts the 
photonic bandgap, thus causing the red shift of the spectra. Similarly, the change of 
the defocusing nonlinear coefficient causes a reduction of the refractive index, 
leading to the blue shift. As concerns potential applications of such reflection
and transmission spectra due to the Kerr nonlinearity, the ability to shift both of them with respect to
the input wavelengths is a vital asset for the use in all-optical signal
processing, including the design of band-selection filters, mode converters,
and wavelength demultiplexers \cite{agrawal2003}.
\begin{figure}[t]
\centering
\includegraphics[width=1\columnwidth]{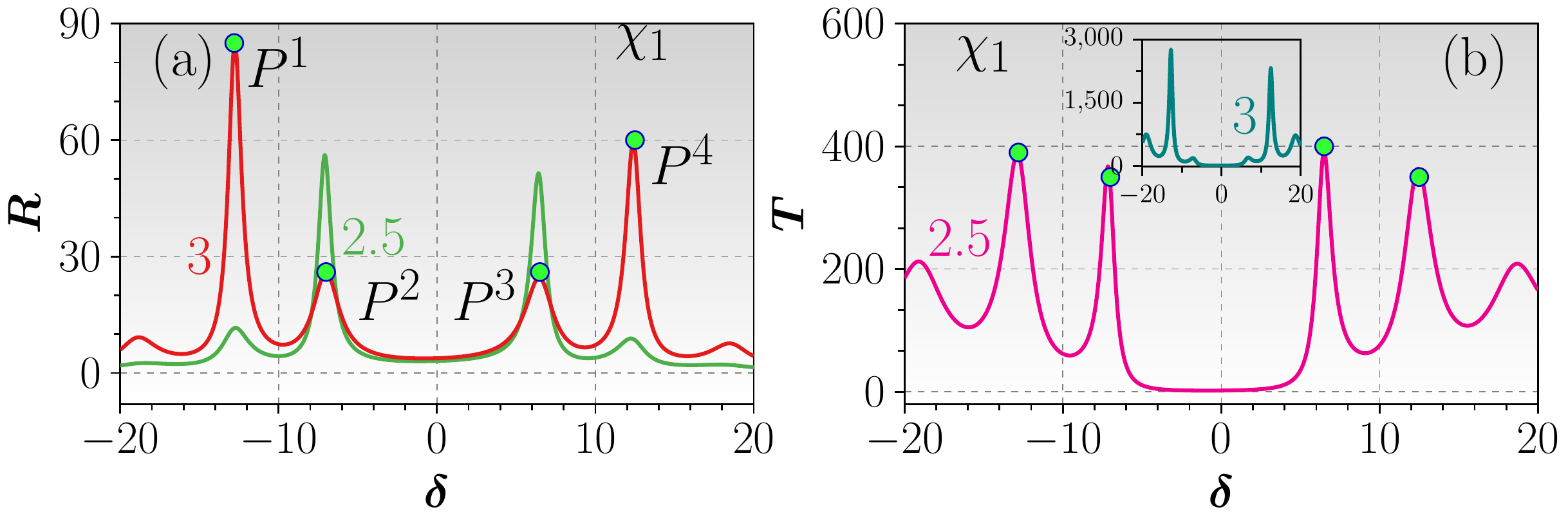}\newline
\includegraphics[width=0.9\columnwidth]{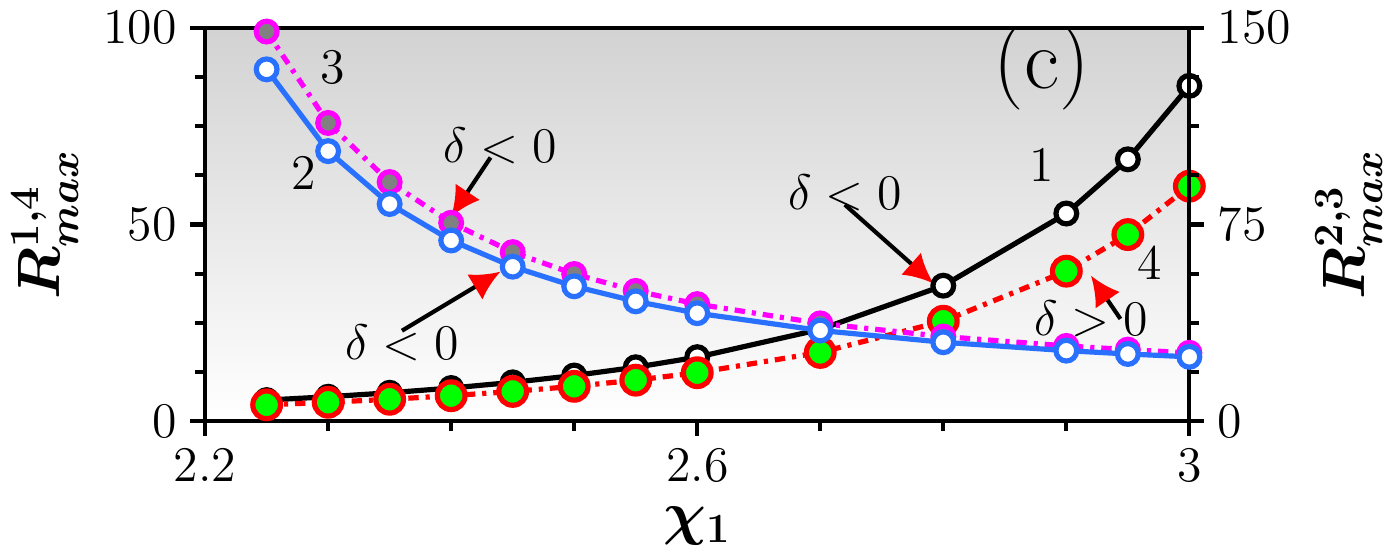}\hspace{-0.1cm} %
\includegraphics[width=0.93\columnwidth]{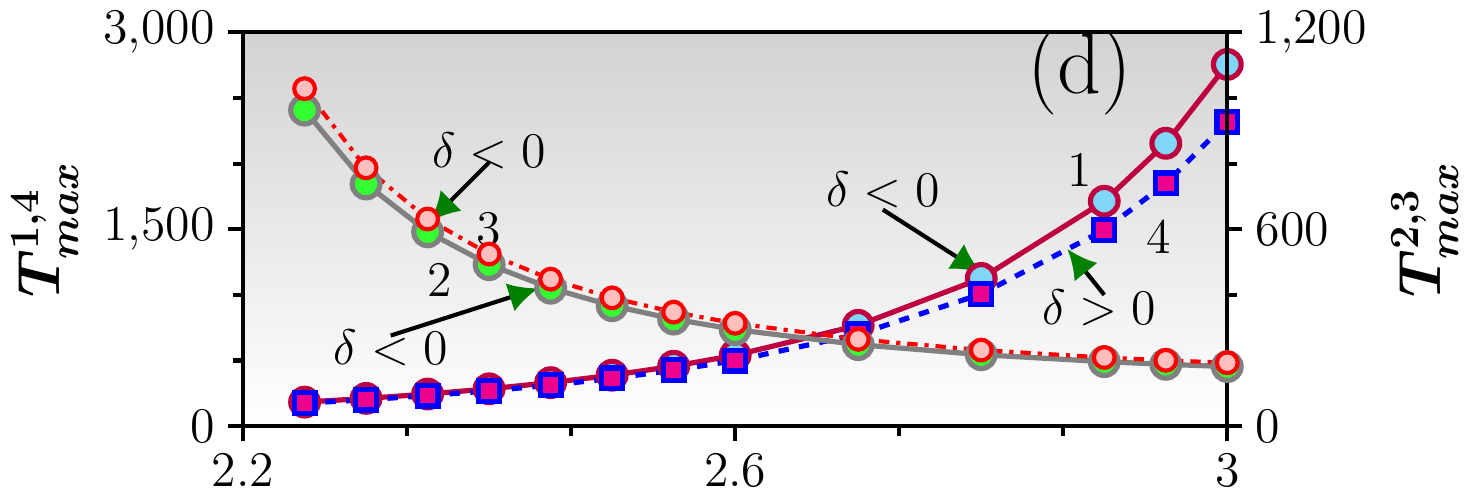}
\caption{(a,b) Dependence of strong secondary spectral resonances of the
reflectivity and transmissivity on the variation of gain in the PRI channel.
The corresponding maximum reflectivity and transmissivity for different
numbers of spectral peaks (denoted by $P^{1,2,3,4}$) are shown in (c,d). The
system's parameters are the same as in Fig. \protect\ref{fig2}.}
\label{fig3}
\end{figure}

To elucidate the detrimental role of the intrinsic loss, $\chi _{2}$, in the
NRI channel, without adding gain to the PRI one, Fig. \ref{fig1}(e)
demonstrates that the corresponding reflection band in the NRI channel
suffers suppression of both its magnitude and width with the increase of $%
\chi _{2}$. On the other hand, the transmission spectra show only marginal
changes in Fig. \ref{fig1}(f), under the variation of $\chi _{2}$. A
noteworthy effect is observed near the edges of the stopband (where zero
transmission takes place), in the form of red shift, while a blue shift
is found outside of the stopband. Thus, a conclusion is that the loss in the
NRI channel of ADC leads to severe suppression of the reflection spectra,
while the characteristic transmission spectra are not strongly affected.

Figure \ref{fig2} displays the calculated reflection and transmission
spectra of the ADC with gain applied in the PRI core, while keeping zero
loss in the NRI one. With gain $\chi _{1}=0.5$, the reflectivity naturally
features a maximum in the stopband in Fig. \ref{fig2}(a), while decaying to zero
outside of it. The increase of the gain to $\chi _{1}=1$ leads to splitting
of the reflectivity maximum in two peaks located near the edges of the stopband.
The transmission spectrum corresponding to Fig. \ref{fig2}(a) is displayed
in Fig. \ref{fig2}(b), which exhibits growing peaks on both sides of the stopband. It further increases rapidly with the increment of $\chi _{1}$, see the
inset in Fig. \ref{fig2}(b). In  both the reflection
and transmission spectra as seen in Figs. \ref{fig2}(c,d), the peaks feature extremely strong amplification at
$\chi _{1}=1.5$ and $1.75$, which is further observed in Figs. \ref{fig2}%
(e,f), where the maxima in $R$ and $T$ for different values of $\chi_1$ are plotted for both $\delta>0$ and $\delta<0$.

When the gain strength in the PRI channel increases beyond $\chi _{1}\approx
2$, remarkable mode-selective amplification occurs at four distinct
wavelengths, two shorter and two longer ones, as seen in Figs. \ref{fig3}%
(a,b). In particular, at $\chi _{1}=2.5$, the two peaks near the edges of the
stopband feature more amplification than those found outside of the
stopband. On the contrary, when $\chi _{1}=3$, the pair of peaks outside the
stopband show giant amplification, while the two others, close to the edges of the
stopband, are suppressed, in comparison with their counterparts at $\chi
_{1}=2.5$. This trend is confirmed by the dependence of the peak
reflectivity on continuous variation of the gain in the PRI channel, as
shown in Fig. \ref{fig3}(c). A similar amplification pattern is exhibited by
the corresponding transmission spectra in Figs. \ref{fig3}(b) and \ref{fig3}%
(d), with a difference that the magnitude of the amplification is much
higher in comparison with the reflectivity.
\begin{figure}[t]
\centering
\includegraphics[width=1\linewidth]{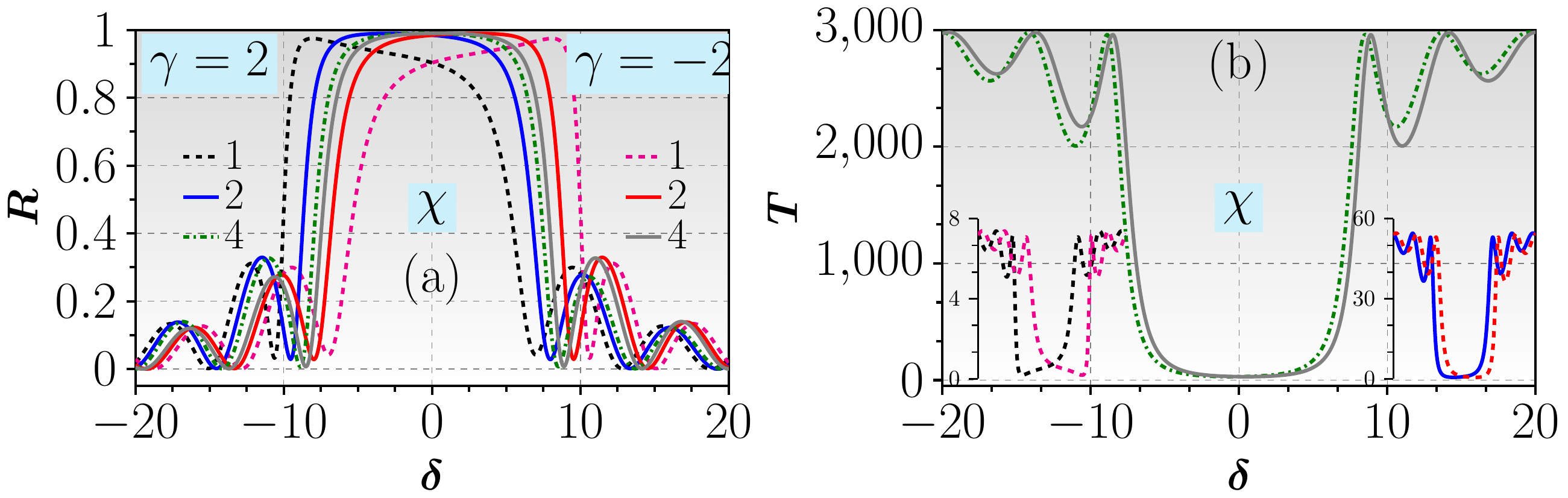}\newline
\includegraphics[width=0.5\linewidth]{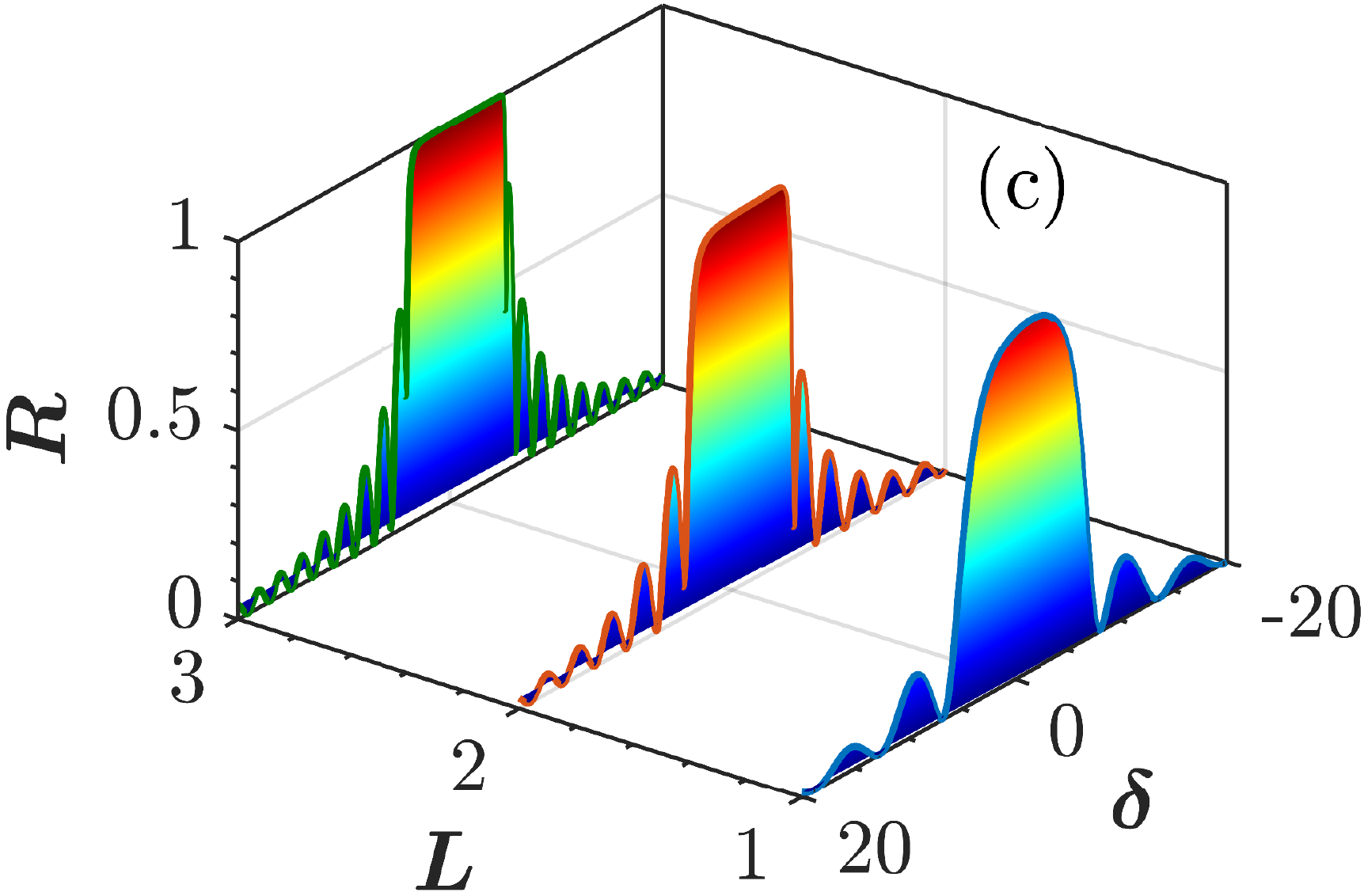}\includegraphics[width=0.5%
\linewidth]{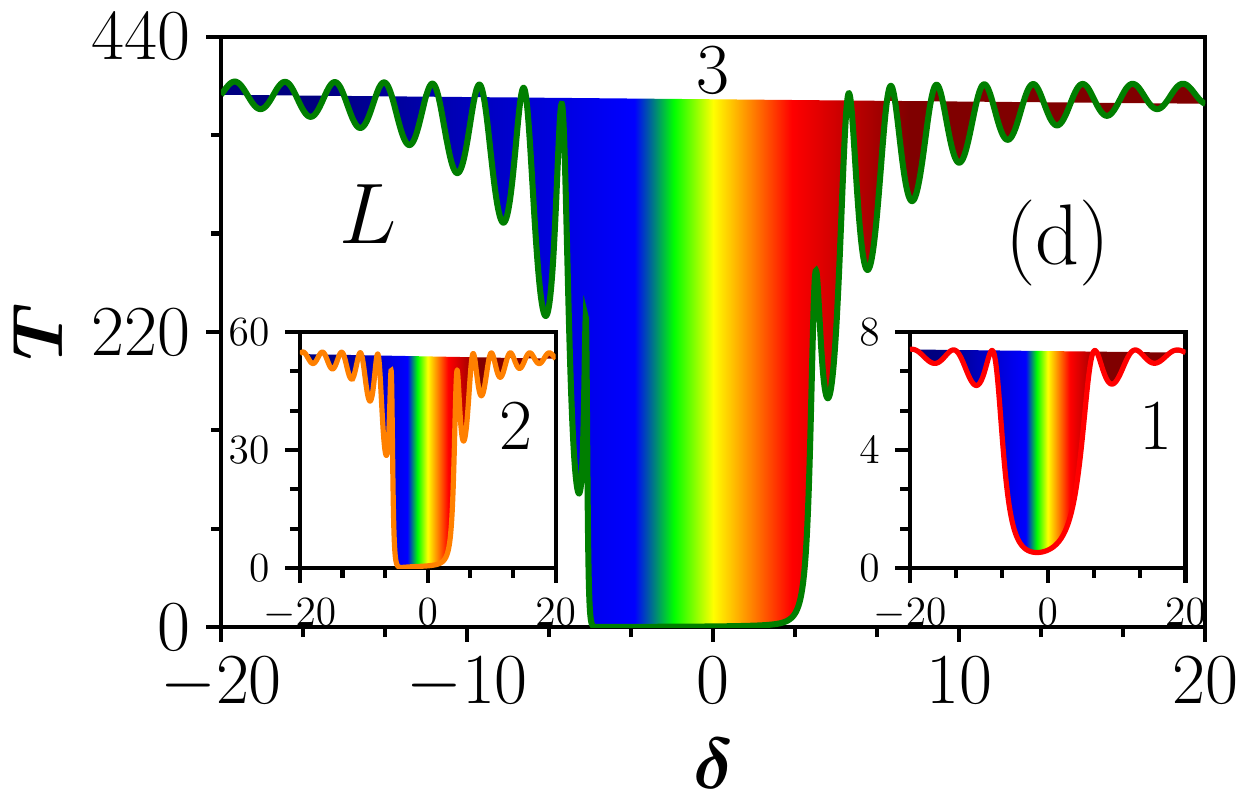}\newline
\includegraphics[width=0.5\linewidth]{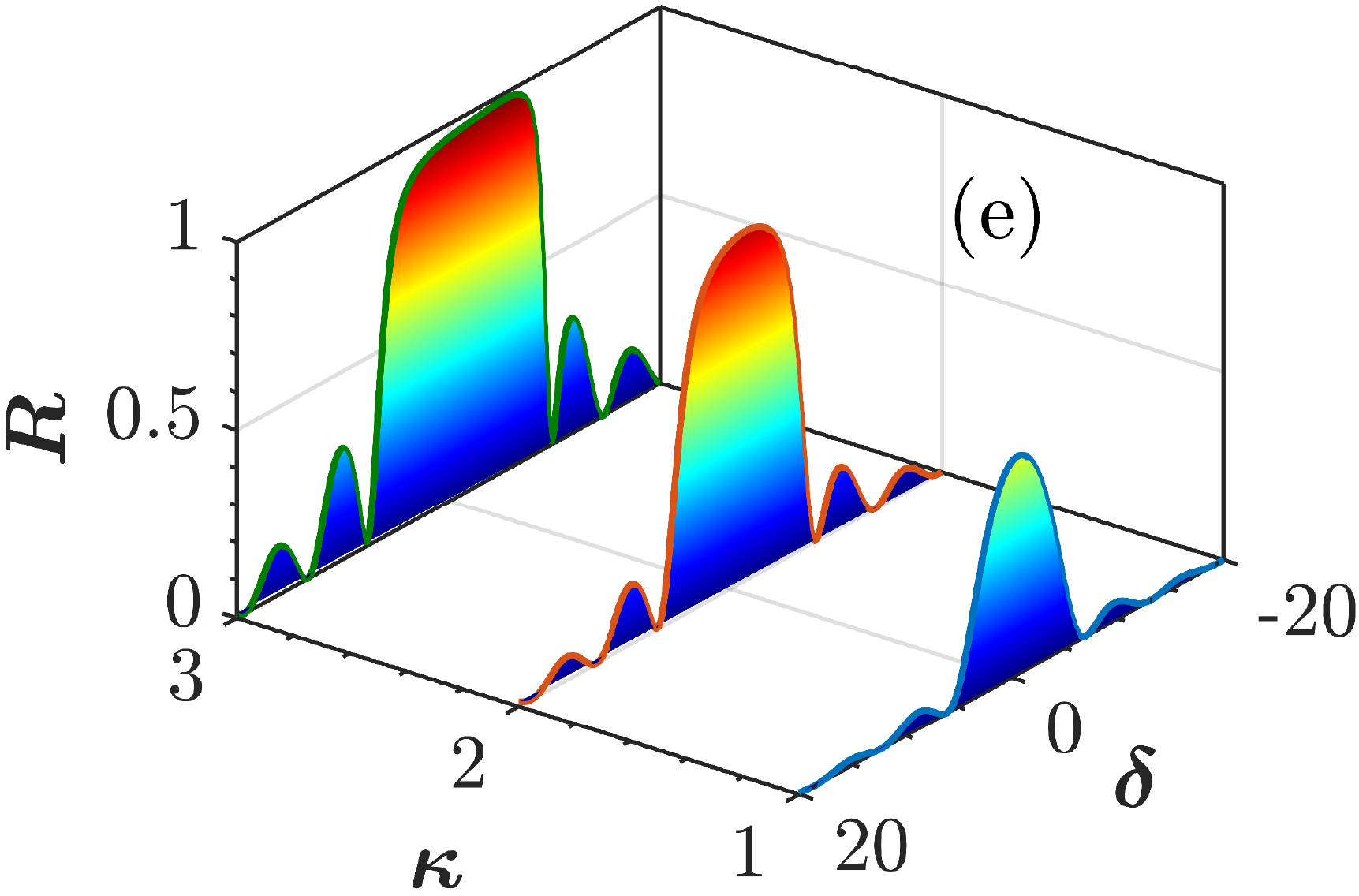}\includegraphics[width=0.5%
\linewidth]{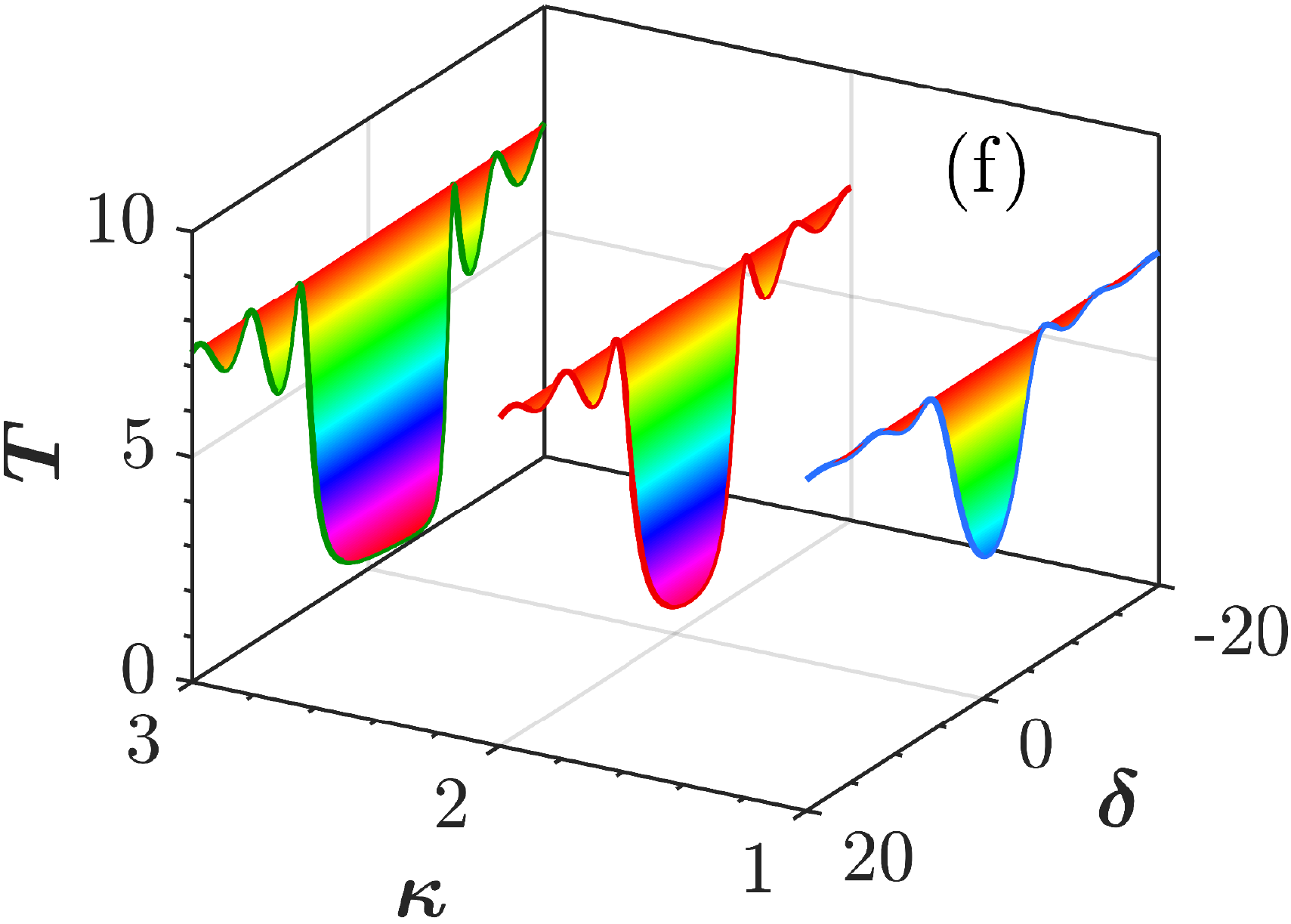} \includegraphics[width=1\linewidth]{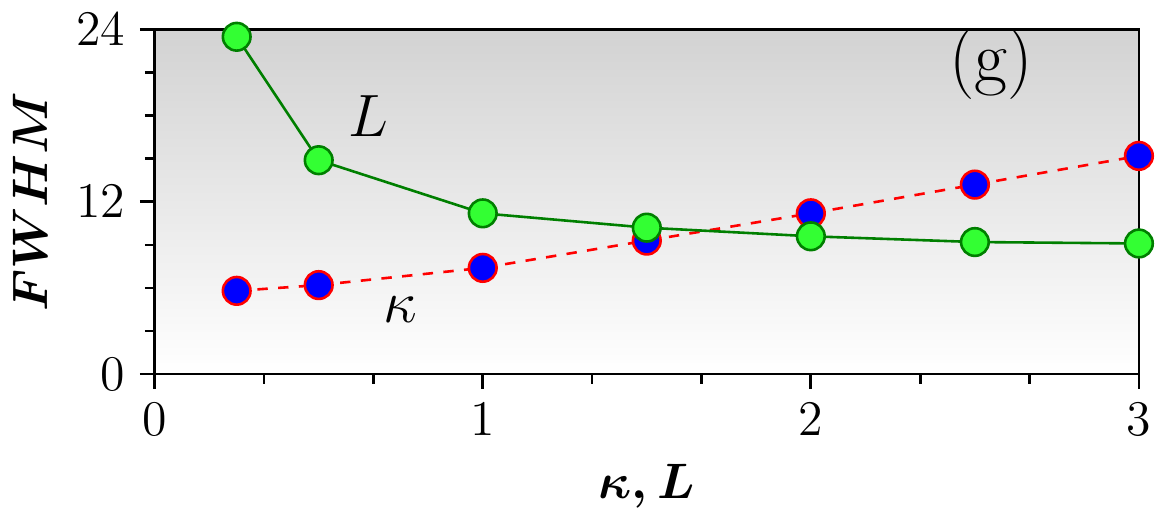}
\caption{(a,b) Nonlinear reflection and transmission spectra as a function
of mismatch $\protect\delta $ for the $\mathcal{PT}$-symmetric ADC, with
equal gain and loss in the PRI and NRI cores ($\protect\chi _{1}=\protect%
\chi _{2}\equiv \protect\chi =1$). Panels (c,d) and (e,f) show,
respectively, the impact of the coupler's length, $L$, and the inter-core
linear coupling, $\protect\kappa $. The bottom panels display the bandwidth
(FWHM) of the reflection spectra versus $\protect\kappa $ and $L$.}
\label{fig4}
\end{figure}

The studies of $\mathcal{PT}$-symmetric systems have drawn a great deal of
interest in couplers with equal magnitudes of gain and loss in two
parallel-coupled cores \cite{Driben}-\cite{Barash}, \cite%
{ReviewPT1,ReviewPT2}, \cite{govindarajan2019}. Results for this case are
collected in Fig. \ref{fig4}. First, it is relevant to stress the effects of
equal gain-loss coefficients on the ADC spectra. As the value of the
gain-loss increases, the spectrum shifts towards shorter or longer
wavelengths if the nonlinearity is, respectively, self-focusing or
defocusing in both the amplified gain and lossy channels, as seen in Fig. %
\ref{fig4}(a). In Fig. \ref{fig4}(b), the transmission spectra exhibit
spectral resonances with huge amplification, without attaining any lasing
behavior, even if the gain is very large, in comparison to the spectra in
the absence of gain and loss. Further, in Fig. \ref{fig4}(a) it is seen that
the increase of the self-focusing coefficient (in particular, to values $%
\gamma _{1}=\gamma _{2}=2$) in ADC with equal gain and loss gives rise to a
stronger red shift in the nonlinear spectra. On the other hand, stronger
self-defocusing nonlinearity drives the blue shift [see red solid and dashed
lines in Figs. \ref{fig4}(a,b)]. In the system including both focusing and
defocusing nonlinearities, the stopband remains flat, completely reflecting
the incident light ($R=1$). This implies keeping ideal spectra, with the
photonic bandgap staying in its original position, the same as in the linear
system (which is not shown here). Even for the transmission spectra in the
PRI channel, conclusions suggested by Fig. \ref{fig4}(a) remain true, as can
be seen in Fig. \ref{fig4}(b). Note that here too, a distinctive feature of
the nonlinear ADC is huge amplification in the transmission spectra, even in
the presence of loss. Thus, the remarkable ability to control the flat broad
stopband and the magnitude of the transmissivity by adjusting the values of the
gain and loss coefficients and wavelength of the input signal makes the ADC
an appropriate element for all-optical signal-processing and demultiplexing
applications.

Finally, Fig. \ref{fig4}(c) shows that the ADC length plays an essential
role in broadening or narrowing the spectral range, as well as in the
enhancement of reflectivity. In particular, when $L=1$ and $\kappa =2$,
the spectral range is broad, but the peak value of the reflectivity is low.
However, when the length increases to $L=2$, the reflectivity in the
stopband increases, while the spectrum shrinks. At $L=3$, the reflectivity
is nearly flat for wavelengths near $\delta =0$, and its magnitude is close
to $1$. The respective transmission spectra exhibit similar peculiarities,
but with transmissivity much higher than the reflectivity at the same
parameters, see Fig. \ref{fig4}(d). Conversely, the region of the nearly
flat spectrum in the stopband can be expanded and made still flatter (and
closer to $1$, as in the case of reflectivity) by increasing the strength of
the inter-core coupling, $\kappa $, as seen in Fig. \ref{fig4}(e). Once
again, the transmission spectra shown in Fig. \ref{fig4}(f) demonstrate a
similar behavior, but with values of the transmissivity much larger than the
corresponding reflectivity (by a factor of $\simeq 8$). In both cases, it is
worthy to note that the bandwidth of spectra decreases (increases) when the
length (inter-core coupling parameter) is varied, as seen in Fig. \ref{fig4}%
(g).

In conclusion, we have shown that ADCs (anti-directional couplers) with the
Kerr nonlinearity in their cores exhibit novel transmission and reflection
spectra, which may be relevant to fundamental studies and potential
applications. The gain applied to one of the two cores (the PRI channel)
maintains the effect of the wavelength-selective amplification, while equal
gain and loss acting in the two cores effectively restore both the
reflection and transmission spectra, which are suppressed by the dominating
loss. The former setting also gives rise to transmission and reflection
spectra with a broad flat stopband, along with inducing red and blue shifts.
These properties suggest that ADC may serve as an essential component in
integrated lightwave data-processing systems. As an extension of the
analysis, it may be relevant to consider a composite waveguide, built of
periodically alternating ADC segments with switched NRI/PRI structure.
Previously, a similar setting composed of segments with periodically
switching gain and loss segments was demonstrated to provide strong
stabilization of $\mathcal{PT}$-symmetric solitons \cite{Driben2}.

\textbf{Funding}: Science and Engineering Research Board (SERB) of India,
PDF/2016/002933 and SB/DF/04/2017). Israel Science Foundation, grant No.
1286/17.

\textbf{Disclosures}. The authors declare no conflicts of interest.

\end{document}